\documentstyle[12pt,epsfig,wrapfig]{article}
\include{times,mathptm}

\renewcommand{\section}[1]{\vspace{6pt} \noindent\mbox{#1} \newline \noindent}
\renewcommand{\subsection}[1]{\vspace{6pt} \noindent\mbox{\underline{#1}} 
\newline \noindent}
\renewcommand{\subsubsection}[1]{\vspace{6pt} \noindent\mbox{\underline{#1}}
\noindent}

\newfont{\sansb}{cmssbx10}
\newfont{\sans}{cmss10}

\begin{document}
{\center \LARGE {\bf UNDERGROUND MUONS IN SUPER-KAMIOKANDE}
\vspace{6pt}\\}
The Super-Kamiokande Collaboration, presented by J. G. Learned$^1$
\vspace{6pt}\\
{\it $^1$Department of Physics and Astronomy, University of Hawaii,
Honolulu, HI USA \vspace{-12pt}\\}

{\center Submitted to 25th ICRC at Durban, South Africa\\}

{\center ABSTRACT\\}

The largest underground neutrino observatory, Super-Kamiokande,
located near Kamioka, Japan has been collecting data since April
1996.  It is located at a depth of roughly 2.7 kmwe in a zinc mine
under a mountain, and has an effective area for detecting
entering-stopping and through-going muons of about $1238~m^2$ for
muons of $>1.7~GeV$.  These events are collected at a rate of 1.5
per day from the lower hemisphere of arrival directions, with 2.5
muons per second in the downgoing direction.

\begin{wrapfigure}{O}{7.5cm}
\epsfig{file=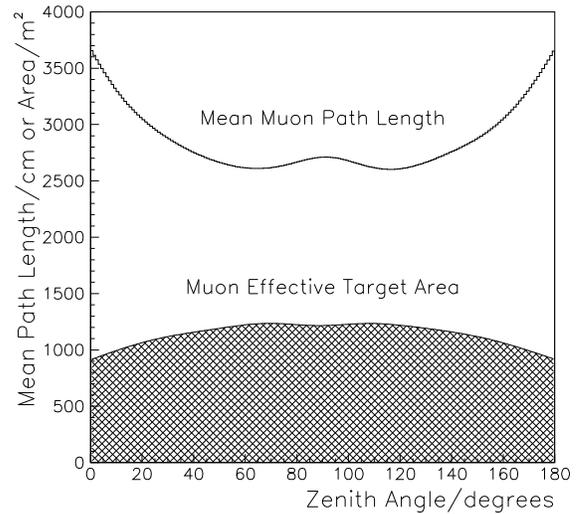,width=7.5cm,height=7.5cm,angle=0}
\caption{\it The effective area for muons versus zenith angle in
Super-Kamiokande. The filled region is for muons with more than 1.7
GeV of energy upon entering the inner detector, including those that
stop in the detector.  The upper curve shows the mean pathlength in
the detector for throughgoing muons.}
\label{fig:area}
\end{wrapfigure} 

We report preliminary results from 229 live days analyzed so far
with respect to zenith angle variation of the upcoming muons. These
results do not yet have enough statistical weight to discriminate
between the favored hypothesis for muon neutrino oscillations and
no-oscillations.  We report on the search for astrophysical sources
of neutrinos and high energy neutrino fluxes from the sun and earth
center, as might arise from WIMP annihilations. None are found.  We
also present a topographical map of the overburden made from the
downgoing muons.  The detector is performing well, and with several
years of data we should be able to make significant progress in this
area.

\setlength{\parindent}{1cm}
\section{THE NEUTRINO DETECTOR}
The Super-Kamiokande detector, located in a mine tunnel in Western
Japan, has been accumulating data for more than one year now, having
begun recording data on 1 April 1996.  The detector is described
elsewhere (Conner, 1997 in these proceedings, and references
therein).  Herein we focus upon the preliminary results obtained
upon upcoming and sidegoing muons produced by cosmic ray muon
neutrino (and anti-neutrino) interactions in the rock surrounding
the detector.  Simply, one can picture the instrument as a 36.2 m tall
by 33.8 m diameter cylindrical Cherenkov inner detector with surrounding
veto counter.

\section{DATA SAMPLE AND REDUCTION}
Data is recorded at about 12 event triggers per second, depending
upon the low energy threshold setting (typically around $5.5~MeV$).
The downgoing cosmic ray muons account for about 2.5/sec, and
constitute the bulk of the recorded data, which total nearly $25
~GBytes/day$.  Solar neutrinos and contained neutrino interactions
occur at the rate of about 19/day and 9/day, respectively.  We find
upgoing entering and throughgoing events at about 1.5/day.  The
typical muon results in 5000 phototube signals within $200~ns$ and
therefore triggering efficiency is not an issue (as it is for low
energy events).

\begin{wrapfigure}{l}{7.5cm}
\epsfig{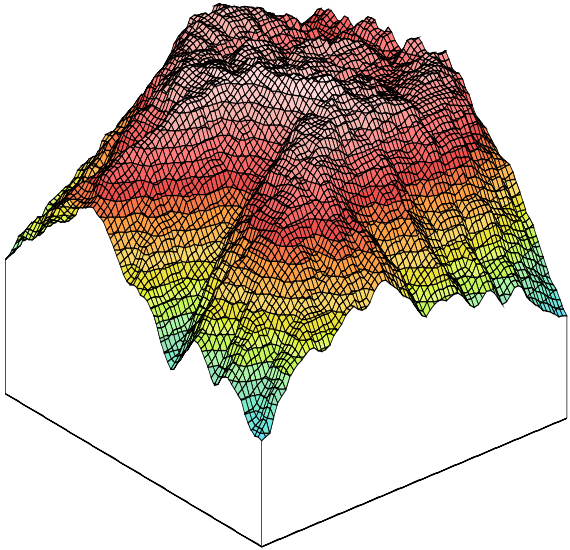}
\caption{\it Topographic maps of the mountain housing the
Super-Kamiokande instrument made with downgoing muon data.
The view is towards the East, from 15 degrees elevation.}
\label{fig:topo_muons}
\end{wrapfigure} 

It is necessary to fit and remember all muon trajectories for the
solar neutrino analysis, wherein a region around each muon track is
rejected for some time after muon traversal in order to eliminate
muon spallation events.  This analysis thus includes the first step
in muon data reduction.  The filtering for upcoming muon events
proceeds to a level of about 100 events per true upcoming muon, and
from there is human scanned.  Fitting of directions is finally done
by hand. Comparison between independent fits and machine fits
indicate that angles are good to $1.3^{\circ}$ and that entry points
are reproduceable to about $0.41~m$, both quite acceptable
values.

As to the efficiency for extracting upcoming events, we find through
independent analysis chains that our finding efficiency for
throughgoing upcoming muons is nearly 100\% for long tracks.  There
is some ambiguity for corner clipping events of a few meters track
length in the detector, and for stopping muons where the contained
track length is difficult to determine.  We thus impose a minimum of
$7~m$ muon track length in the detector for acceeptance into the
data sample.

\begin{wrapfigure}{r}{7.5cm}
\epsfig{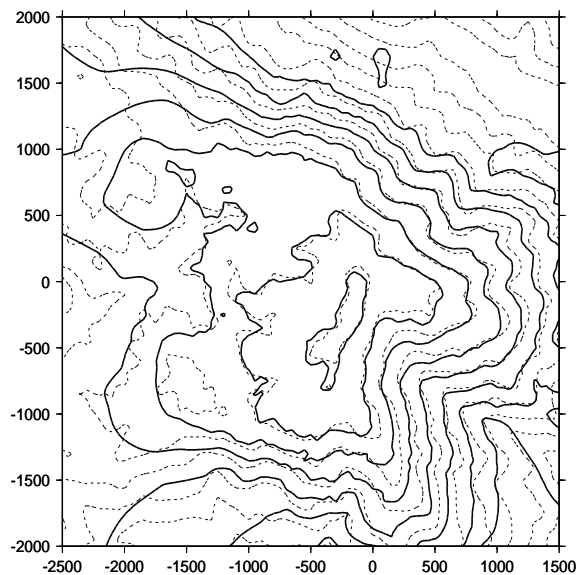}
\caption{\it Topographic map calculated from the observed muon flux
at the Super-Kamiokande detector superposed on the geodetic map
(dashed contours).  The contours are at 100 m intervals, the scale
is in m, and the top is SE.}
\label{fig:superimpose_map}
\end{wrapfigure} 

Due to the predominance of numbers of downgoing muons, and the fact
that the mountain has relatively thin areas (see Figures
\ref{fig:topo_muons} and \ref{fig:superimpose_map}), there are many
muons out to zenith angles of about $85^{\circ}$.  We have cut the
data sample at the horizon.  There still may be some contamination
from back scattered events arriving from a few degrees below the
horizon, but we estimate this to be less than 2 events in our
present data sample, a negligible effect.

\section{RESULTS}
We have analyzed so far 229 live days of data, with a net livetime
fraction of roughly 80\%, including experiment run-in.  In this
sample we have 267 muons which both enter and exit the inner
detector, and which have more than $7~m$ of path length. The
effective flux, averaged over the lower hemisphere is $1.76\pm 0.10
\times 10^{-13} muons/cm^2/sec/sr$ (statistical error only). This
may be compared to predicted fluxes of 1.99 (Agrawal 1996 and Gluck
1996) and 1.86 (Honda, 1995 and Gluck, 1995), in the same units.

Note that, as illustrated in Figure \ref{fig:area}, while the
minimum muon energy is $1.7~GeV$, the effective energy threshold is
near to $6~GeV$.  We are also analyzing the entering-stopping muons,
which are a substantial fraction of the throughgoing events, but
this analysis is not ready for presentation yet.
\begin{wrapfigure}{l}{7.5cm}
\epsfig{file=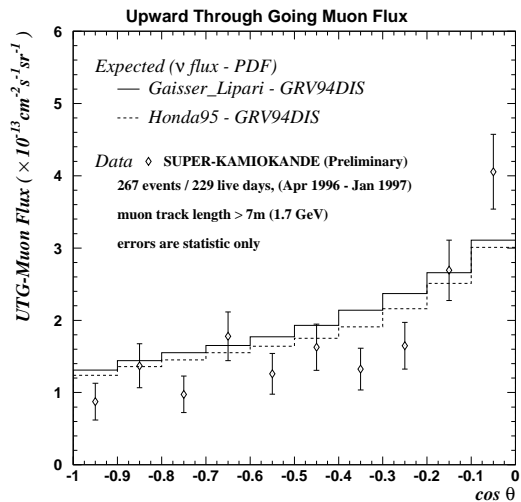,width=7.5cm,height=7.5cm,angle=0}
\caption{\it The zenith angle distribution of through-going upcoming
muon events with at least 7 m pathlength in Super-Kamiokande, from
the first 200 day data sample.}
\label{fig:zendist}
\end{wrapfigure} 

\subsection{Muon Tomography}
As an analysis exercise and a check on our detector survey, we have
used the measured directions of downgoing muons to generate a
tomographic projection of the mountain overhead.  For this purpose
we utilized a sample of $2 \times 10^7$ fitted muon trajectories and
a simple approximation to the attenuation due to a constant density
overburden.  The resulting muon contour map was fitted to the
geodetic survey data with parameters of detector position, azimuth,
depth, and overburden mean density.  An oblique view of the mountain
made from muon data is show in Figure \ref{fig:topo_muons}, and a
comparison of the muons map and the geodetic survey are presented in
Figure \ref{fig:superimpose_map}.

\begin{wrapfigure}{r}{12.5cm}
\epsfig{file=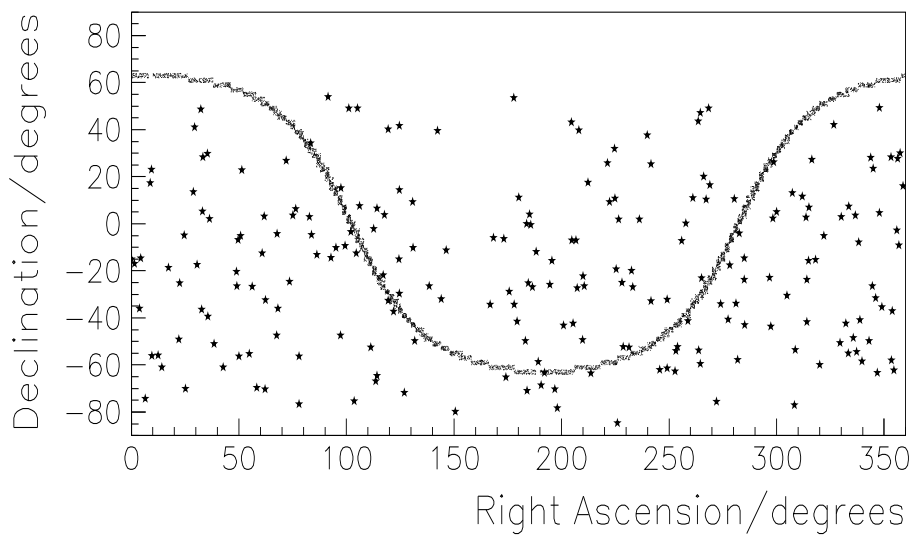,width=12.5cm,height=7.5cm,angle=0}
\caption{\it The distribution of the upcoming muon directions in the
first 200 day sample from Super-Kamiokande, plotted in Right
Ascension and Declination. The line shows the galactic plane.}
\label{fig:ra-dec}
\end{wrapfigure}

\subsection{Atmospheric Neutrino Angular Distribution}
In Figure \ref{fig:zendist} we show the angular distribution of
upcoming muons from this initial 229 day sample.  Predicted angular
distributions as indicated above, are superposed. The limited
statistics prevent drawing of any conclusions with regard to
neutrino oscillations as yet.

\subsection{Sky Map}

In Figure \ref{fig:ra-dec} we present the initial sky map of event
arrival directions of upcoming neutrino events in Super-Kamiokande.
The plane of the galaxy is superposed upon the Figure.  We have
tested the data with the twopoint correlation function and several
other tests for non-uniformity upon the sky and find nothing
indicating extra-terrestrial neutrino point sources.  The
distribution of neutrino directions on the sky is consistent with
isotropy.

\subsection{Events in the Direction of the Sun}
\begin{wrapfigure}{l}{7.5cm}
\epsfig{file=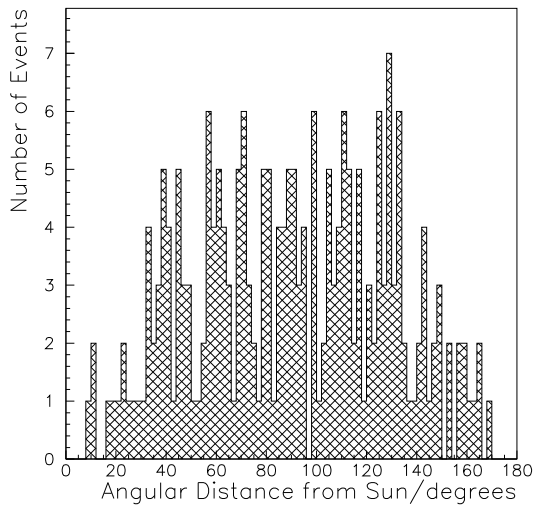,width=7.5cm,height=7.5cm,angle=0}
\caption{\it The distribution of 219 upcoming muon directions versus
angular distance from the sun. No events are seen to be emanating
from the solar direction.}
\label{fig:soldist}
\end{wrapfigure}

In Figure \ref{fig:soldist}, we present the initial angular
distribution of neutrino events origin relative to the sun's
direction in the sky.  One sees no events pointing towards the sun,
and this can be interpreted into limits on WIMP annihilations.  As
evidenced in Figure \ref{fig:zendist} there is also no excess of
events emanating from the earth's core.  Again, because of the as
yet limited statistics, we do not push the existing limits, but do
confirm their legitimacy.

\section{CONCLUSIONS}
The Super-Kamiokande detector is now fully in operation, and the
data analysis is well underway.  While the Super-Kamiokande is vastly
more massive than predecessor solar neutrino experiments, such that
it overwhelmed previous experiments statistically within one month
of operation, Super-Kamiokande is only about a factor of three
larger in muon area than the previous largest detector.  It will be
a few years before Super-Kamiokande dominates the event total.
However, because of the high angular resolution and the ability to
set a higher energy threshold (by muon range), this data will begin
to break new ground within about one year.  So far we see nothing
unexpected, confirming previous results.  We will, of course,
continue to monitor for new phenomena, temporal correlations with
gamma ray bursts and will search for ultra high energy events
traversing the detector.  The detector is functioning well and we
look forward to many years of data analysis.

\section{ACKNOWLEDGEMENTS}
Many colleagues contributed to this effort, in particular those of
the muon analysis group of the Super-Kamiokande Collaboration in
Japan and the U.S., and more generally of the entire Collaboration
who built and operate the Super-Kamiokande instrument.  We thank
Todor Stanev for data and computer code used in calculations of
neutrino fluxes and the probability for neutrino conversion to
muons.  We acknowledge the support of the Japanese Ministry of
Education and the U.S. Department of Energy.

\section{REFERENCES}
\setlength{\parindent}{-5mm}
\begin{list}{}{\topsep 0pt \partopsep 0pt \itemsep 0pt \leftmargin 5mm
\parsep 0pt \itemindent -5mm}
\vspace{-15pt}

\item Conner, Z., et al., Proc. 25th ICRC, paper HE 4.1.22 (1997).

\item Agrawal, V., T.K.~Gaisser, Paolo Lipari and Todor Stanev,
Phys. Rev. D{\bf 53}, 1314 (1996).

\item Gluck, M., E.~Reya, A.~Vogt, Z. Phys. C {\bf 67}, 433 (1995).

\item Honda, M., T.~Kajita, K.~Kasahara, and S.~Midorikawa,
hep-ph/9511223.

\end{list}

\end{document}